\documentclass[preprint,authoryear,12pt]{elsarticle}

\usepackage{graphicx}
\usepackage{amsmath,amssymb}
\usepackage{dsfont}
\usepackage{color}

\newcommand{\eps}{\varepsilon}
\renewcommand{\leq}{\leqslant}
\renewcommand{\geq}{\geqslant}
\renewcommand{\hat}{\widehat}
\renewcommand{\tilde}{\widetilde}
\newcommand{\cqfd}{\hfill{$\Box$}}
\newcommand{\ds}{\displaystyle}
\newcommand{\N}{\mathbb{N}}
\renewcommand{\P}{\mathbb{P}}
\newcommand{\R}{\mathbb{R}}

\newtheorem{theo}{Theorem}

\newtheorem{lem}{Lemma}

\newcommand{\Mpt}{\widetilde{s}_{n}}
\newcommand{\mpt}{\widetilde{r}_{n}}
\newcommand{\Mph}{\widehat{s}_{n}}
\newcommand{\qt}{\bar{q}}
\newcommand{\Qt}{\bar{Q}}

\begin{document}

\begin{frontmatter}



\title{Estimation of a convex discrete distribution}


\author[CD]{C\'ecile Durot}
\author[FK1,FK2]{Fran\c cois Koladjo}
\author[FK1]{Sylvie Huet\corref{cor1}}
\author[SR1,SR2]{St\'ephane Robin}

\address[CD]{UFR SEGMI, Universit\'e Paris Ouest Nanterre La
  D\'efense, F-92001, Nanterre, France}
\address[SH1]{UR341 MIA, INRA, F-78350 Jouy-en-Josas, France}
\address[FK2]{CIPMA-Chaire UNESCO, FAST, UAC, 072BP50 Cotonou, B\'enin}
\address[SR1]{UMR518 MIA, INRA F-75005 Paris, France}
\address[SR2]{UMR518 MIA, AgroParisTech F-75005 Paris, France}
\cortext[cor1]{Corresponding author.\\{\sl E-mail address:} sylvie.huet@jouy.inra.fr}

\begin{abstract}
Non-parametric estimation of a convex discrete distribution may be of
 interest in several applications, such as the estimation
 of species abundance distribution in ecology. In this paper we study
 the least squares estimator of a discrete  distribution under the
 constraint of convexity. We show that this estimator exists and is
 unique, and that  it always outperforms the classical empirical 
estimator in terms of the $\ell_{2}$-distance. We provide an algorithm
 for its computation, based on the support reduction algorithm. We compare
 its performance to those of the empirical estimator, on the basis of a
 simulation study.

\end{abstract}


\end{frontmatter}

% \linenumbers

\section{Introduction} \label{intro.st}

The nonparametric estimation, based on the observation of $n$
i.i.d. copies $X_{1}$,~\dots, $X_{n}$, of the distribution of a
continuous random variable under a monotonicity constraint, has
received a great deal of attention in 
the past decades, see \cite{BW06} for a review. The most studied
constraint is the monotonicity of the density function. It is
well-known that the nonparametric maximum likelihood estimator of a
decreasing density function 
over $[0,\infty)$ is the Grenander estimator defined as the left-continuous
slope of the least concave majorant of the empirical distribution
function of $X_{1}$,~\dots, $X_{n}$. This estimator can be  easily
implemented using the PAVA (pool adjacent violators algorithm) or a
similar device, see \cite{BBBB}. 
Another well studied constraint is the monotonicity of the first
derivative of the density, such that the density function is assumed to
be convex (or concave) over a given interval. It was shown by
\cite{GJW01} that both the least squares estimator and the
nonparametric maximum likelihood estimator under the convexity
constraint exist and are unique. However, although a precise
characterisation of these estimators is given in that paper, their
practical implementation is a non-trivial issue: it requires
sophisticated iterative algorithms that use a mixture
representation, such as the {\it support reduction algorithm} described
in \cite{GJW08}. The nonparametric maximum likelihood of a log-concave
density function (i.e., a density function $f$ such that $\log(f)$
is a concave function) was introduced in \cite{R06} and algorithmic
aspects were treated in \cite{R07} and in \cite{DHR07}, where an
algorithm similar to the support reduction algorithm is defined. 

Recently, the problem of estimating a discrete probability mass
function under a monotonicity constraint has attracted attention:
\cite{JaW09} considered the non-parametric 
estimation of a monotone distribution and \cite{BJR11} considered the
case of a log-concave distribution. 

In this paper, we consider the
nonparametric estimation of a discrete distribution on $\N$ under the convexity
constraint. This problem has not yet been considered in the
literature, although it has several applications, such as the estimation
of species abundance distribution in ecology. In this field,
the terms ``nonparametric methods'' often refer to finite mixtures of parametric
distributions where only the mixing distribution is inferred in a
nonparametric way, see e.g. (\cite{BoK06}, \cite{BDK05-StatMethAppli},
\cite{ChS04}).

We study the least squares estimator of a discrete
distribution on $\N$ under the constraint of convexity. 
First, we prove that this estimator exists and is unique, and that it
always outperforms the classical empirical estimator in terms of the
$\ell_{2}$-distance.  Then, we consider computational issues. 
Similar to the continuous case, we prove that
a representation of convex discrete
distributions can be given in terms of  a --~possibly infinite~--
mixture of triangular
functions on ${\mathbb N}$, and, based on this characterization, we derive an algorithm that
provides the least squares estimate, although both the number of components
in the mixture and the support of the estimator are unknown. This algorithm is an adaptation to our problem of the support reduction algorithm in \cite{GJW08}. Finally, we 
assess the performance of the least squares estimator under the convexity
constraint through a simulation study.

The paper is organized as follows. Theoretical properties of the constrained least squares estimator are given in Section~\ref{sectionLSE}. Section \ref{TriBasisAlgo.st} is devoted to computational issues. A similation study is reported in Section~\ref{simul.st}, and the proofs are postponed to Section~\ref{Proofs.st}.

%In Section~\ref{sectionLSE} we  present the main results on the existence and uniqueness of the least squares estimator of a discrete distribution under the constraint of convexity. Then we show that it always outperforms the classical empirical estimator in terms of the $\ell_{2}$-distance. In Section~\ref{TriBasisAlgo.st} we derive the decomposition of discrete convex functions on the set of triangular functions on ${\mathbb N}$ and use it to introduce an algorithm for solving the minimization procedure based on this representation. In Section~\ref{simul.st}  we  assess the performances of the least squares estimator under convexity constraints through a simulation study. The proofs are postponed to Section~\ref{Proofs.st}.

\paragraph{Notation.}
Let us define some notation that will be used throughout the paper
\begin{itemize}

\item ${\cal K}$ is the set of convex functions $f$ on ${\mathbb N}$
  such that $\lim_{i\to\infty}f(i)=0$.  We recall that a discrete
  function $f:{\mathbb N}\to\R$ is convex if and only if it satisfies
  \begin{equation}\label{def: convex(taux)}
    f(i)-f(j)\geq (i-j)\big(f(j+1)-f(j)\big)
  \end{equation}
  for all $i$ and $j$ in ${\mathbb N}$, or equivalently, if and only
  if
  \begin{equation}\label{def: convex(slope)}
    f(i)-f(i-1)\leq f(i+1)-f(i)
  \end{equation}
  for all $i\geq 1$. In particular, any $f\in{\cal K}$ has to be non-negative,
  non-increasing and strictly decreasing on its support. 
  
\item ${\cal C}$ is the set of all convex probability mass functions
  on ${\mathbb N}$, i.e., the set of functions $f \in {\cal K}$
  satisfying $\sum_{i\geq 0}f(i)=1$.
\end{itemize}

\section{The constrained LSE of a convex discrete distribution} \label{sectionLSE}

\subsection{The main result}\label{mainresult.st}

Suppose that we observe $n$ i.i.d. random variables $X_1,\dots,X_n$
that take values in $\N$, and that the common probability mass
function $p_0$ of these variables is convex on $\N$ with an unknown
support. Based on these observations, we aim to build an estimator of
$p_0$ that satisfies the convexity constraint.

For this task, define the empirical estimator $\tilde p_n$ of $p_0$ by
$$\tilde p_n(j)=\frac{1}{n} \sum_{i=1}^n I_{(X_i=j)}$$
for all $j\in\N$, and consider the criterion function
$$Q_n(f)=\frac{1}{2}\sum_{i\geq 0}f^2(i)-\sum_{i\geq 0}f(i)\tilde p_n(i)$$
for all functions $f:\N\to\R$.  The empirical estimator $\tilde p_n$
may be non-convex so in order to build a convex estimator, we minimize
the criterion function $Q_n$ over the set ${\cal C}$. The minimizer
(which exists according to Theorem \ref{theo: LSE} below) is called
the constrained least squares estimator (LSE) of $p_0$ because it also
minimizes the least squares criterion
$$\frac{1}{2}\sum_{i\geq 0}\big(f(i)-\tilde p_n(i)\big)^2
=Q_n(f)+\frac{1}{2}\sum_{i\geq 0}\tilde p_n^2(i).$$
It is clear that in the case where $\tilde p_n$
is convex, the constrained LSE coincides with $\tilde p_n$. On the
other hand, in the case where $\tilde p_n$ is non-convex, the
constrained LSE outperforms  the empirical estimator $\tilde p_n$, as
detailed in Section \ref{sec: empir/constaint}. 

The existence and uniqueness of the constrained LSE of $p_{0}$ over ${\cal C}$ is shown in the
following theorem. It is proved that $\widehat{p}_{n}$ is the
minimizer of $Q_{n}$ over the set ${\mathcal K}$, and has a finite
support. We will denote by $\Mph$, respectively $\Mpt$,
the maximum of the support of $\widehat{p}_{n}$, respectively
$\widetilde{p}_{n}$.

\begin{theo}\label{theo: LSE}
There exists a unique $\hat p_n\in{\cal C}$ such that
$$ Q_n(\hat p_n)=\inf_{p\in{\cal C}}Q_n(p) = \inf_{p\in{\cal
    K}}Q_n(p).$$
Moreover, the support of $\hat p_n$ is finite, and $\Mph
    \geq \Mpt$.
\end{theo}

\subsection{Comparison between constrained and unconstrained estimators}\label{sec: empir/constaint}

In Theorem~\ref{theo:
  empir/constraint}, we show the benefits of using the constrained LSE 
rather than the (unconstrained) empirical estimator $\tilde
p_n$, in
terms of the $l_{2}$-loss. Specifically, the constrained LSE is closer to the
unknown underlying distribution $p_0$ than is the unconstrained
estimator $\tilde p_n$. Moreover, we prove that this happens with a
strictly positive probability (and even, a probability of at least
1/2) whenever $p_0$ is not strictly convex on its support.

\begin{theo}\label{theo: empir/constraint}
Let $p_{0}$, $\widetilde{p}_{n}$ and $\widehat{p}_{n}$ be defined as
in Section~\ref{mainresult.st}. We have the following results:
\begin{equation}
 \sum_{j\geq 0}\big(p_{0}(j)-\hat p_n(j)\big)^2\leq \sum_{j\geq
 0}\big(p_{0}(j)-\tilde p_n(j)\big)^2,  \label{eq.theo2}
\end{equation}
with a strict inequality if $\tilde p_n$ is non-convex.
Assume that there exist $i,j\in\N$ such that $j\geq i+2$, $p_0(i)>0$, and $p_0$ is linear over $\{i,\dots,j\}$. Then,
\begin{equation}\label{eq: tilde p non-convex}
\liminf_{n\to\infty}\P\Big(\tilde p_n\text{ is non-convex}\Big)\geq1/2 ,
\end{equation}
and 
\begin{equation}\label{liminf.eq}\liminf_{n\to\infty}\P\left(\sum_{j\geq 0}\big(p_0(j)-\hat p_n(j)\big)^2< \sum_{j\geq 0}\big(p_0(j)-\tilde p_n(j)\big)^2\right)\geq1/2
\end{equation}
\end{theo}

\paragraph{Remark:} as we shall see in the proof of
Theorem~\ref{theo: empir/constraint}, Equation~(\ref{eq.theo2}) also
holds with $p_{0}$ replaced by  any  $q \in {\mathcal K}$ that belongs to $l_2$,  i.e., that
satisfies $\sum_{j} q^{2}(j) < \infty$.

Now, we consider the estimation of some characteristics of the
distribution $p_{0}$, namely the expectation, the centered moments and
the probability at 0.  As estimators for these characteristics, we
naturally consider similar caracteristics of the constrained and the
unconstrained estimators. Theorem~\ref{moments.th} states that  the
distributions 
$\widetilde{p}_{n}$ and $\widehat{p}_{n}$ have the same expectation, but
the centered moments of the  distribution $\widetilde{p}_{n}$ are lower than those of the
  distribution   $\widehat{p}_{n}$. In particular,  the variance of the distribution of $\widehat{p}_{n}$  is greater than the variance
of $\widetilde{p}_{n}$. Moreover, the constrained estimator $\hat
p_{n}(0)$ is greater than or equal to the unconstrained estimator
$\tilde p_{n}(0)$.  The performance of  $\hat p_{n}$ is compared with
that of $\tilde p_{n}(0)$ through simulation studies in Section \ref{simul.st}.
   
\begin{theo}\label{moments.th}
Let $\widetilde{p}_{n}$ and $\widehat{p}_{n}$ be defined as
in Section~\ref{mainresult.st}. We have for all $u \geq 1$, and
$0 \leq a \leq \widehat{s}_{n}$
\begin{equation}
 \sum_{i=1}^{\widetilde{s}_{n}} |i-a|^{u} \widetilde{p}_{n}(i) \leq
\sum_{i=1}^{\widehat{s}_{n}} |i-a|^{u} \widehat{p}_{n}(i).
\label{moments.ineq}
\end{equation}
Moreover, $\sum_{i=1}^{\widetilde{s}_{n}} i
\widetilde{p}_{n}(i)=\sum_{i=1}^{\widehat{s}_{n}} i
\widehat{p}_{n}(i)$ and $\hat p_{n}(0)\geq \tilde p_{n}(0)$.
\end{theo}
It can be shown that similar results hold for constraint estimators of
a convex density function, where $\tilde p_{n}$ is replaced by an
unconstrained estimator of the density function and $\hat p_{n}$ is
replaced by the corresponding constrained estimator. 
On the contrary, in the case of discrete log-concave distribution, it is shown by ~\cite{BJR11}, see their Equations~(3.5) and ~(3.6), that the
moments of the constrained
 maximum likelihood estimator distribution are smaller than those of the
empirical distribution. These authors refer to similar results  for the maximum
likelihood estimator  of a continuous log-concave density. 

\section{Implementing the constrained LSE}\label{TriBasisAlgo.st}

\subsection{More on convex discrete functions}

The aim of this section is to prove that any $f\in{\cal K}$ is a
combination of the triangular functions $T_j$ defined below, and that
the combination is unique. This compares with Propositions 2.1 and 2.2
in \cite{BW06}, which deals with the case of convex (and more
generally, $k$-monotone) density functions on $(0,\infty)$. For every
integer $j\geq 1$, we define the $j$-th triangular function $T_j$ on
$\N$ by 
\[T_j(i)=\begin{cases}
          \ds\frac{2(j-i)}{j(j+1)}\text{ for all }i\in\{0,\dots,j-1\}\\
           0  \text{ for all integers } i\geq j.
         \end{cases}\]
It should be noticed that $T_j$ is normalized in such a way that it is a probability mass function, i.e., $T_j(i)\geq 0$ for all $i$ and 
$$\sum_{i\geq 0}T_j(i)=1.$$
Moreover, $T_j$ is monotone non-increasing and convex on
$\N$. Hereafter, we denote by ${\cal M}$ the convex cone of
non-negative measures on $\N\backslash\{0\}$.  We denote by $\pi_{j}$,
for $j\in \N\backslash\{0\}$, the components of $\pi \in {\cal M}$.

\begin{theo}\label{theo: mixture k}
Let $f:\N\to [0,\infty)$ such that $\lim_{i\to\infty}f(i)=0.$
\begin{enumerate}
 \item We have $f\in{\cal K}$ if and only if there exists $\pi\in{\cal M}$ such that 
\begin{equation}\label{eq: mixture k}
f(i)=\sum_{j\geq i+1} \pi_jT_j(i)\text{ for all }i\geq 0. 
\end{equation}
\item Assume $f\in{\cal K}$. Then, $\pi$ in (\ref{eq: mixture k}) is uniquely defined by
\begin{equation}\label{eq: pi/p}
\pi_j=\frac{j(j+1)}{2}\big( f(j+1)+f(j-1)-2f(j)\big)\text{ for all }j\geq 1.  
\end{equation}
\item Assume $f\in{\cal K}$. Then, $\pi$ is a probability measure over $\N\backslash\{0\}$ if and only if $f$ is a probability mass function.
\end{enumerate}
\end{theo}

Let us note that according to \eqref{eq: pi/p}, $\pi$ puts mass
  at point $j$ if, and only if, $f$ changes of slope at point
  $j$. Moreover, denoting by $s$ the maximum of the support of $f$ in
  the case where this support is not empty, we see that the greatest
  point where $f$ changes of slope is $s+1$, since the left-hand slope
  of $f$ at this point, $f(s+1)-f(s)$, is strictly negative whereas the right-hand
  slope,  $f(s+2)-f(s+1)$, is zero. Therefore, in the case where the support of $f$ is
  not empty, the greatest point where $\pi$ puts mass is
  $s+1$. Obviously, in case $f(j)=0$ for all $j\geq 0$, we also have
  $\pi_j=0$ for all $j\geq 1$.

\subsection{Algorithm}
Define the criterion function
$$\Psi_n(\pi)=\frac{1}{2}\sum_{i\geq 0}\left(\sum_{j\geq i+1}\pi_jT_j(i)\right)^2-\sum_{i\geq 0}\tilde p_n(i)\sum_{j\geq i+1}\pi_jT_j(i)$$
for all $\pi\in{\cal M}$. The reason why we define such a criterion
function is that 
$\Psi_n(\pi)=Q_n(p)$
for all $p\in{\cal K}$ and $\pi\in{\cal M}$ satisfying (\ref{eq:
  mixture k}) with $f$ replaced by $p$. 
 The constrained LSE of $p_0$ is the unique minimizer of
$Q_n(p)$ over $p\in{\cal K}$. It follows from Theorem \ref{theo:
  mixture k} that there exists a unique $\hat \pi_n\in{\cal M}$ that
minimizes $\Psi_n(\pi)$ over $\pi\in{\cal M}$, and  $\hat p_n$ and
$\hat \pi_n$ are linked by the relation 
\begin{equation}\label{eq: ppihat}
\hat p_n(i)=\sum_{j\geq i+1} \hat \pi_{nj}T_j(i)\text{ for all }i\geq 0.
\end{equation}
Therefore, computing the constrained LSE $\hat p_n$ of $p_{0}$ comes
to computing the measure $\hat\pi_n$ that minimizes $\Psi_n(\pi)$ over
$\pi\in{\cal M}$. Moreover, we know  from Theorems~\ref{theo: LSE} and~\ref{theo:
  mixture k} that  $\hat \pi_n$ is a probability measure and that its
  support is finite with the greatest point %, the maximum of its support being 
  equal to
  $\Mph+1$.

For all $L \geq 1$, let ${\mathcal M}^{L}$ be the set of measures $\pi
\in {\mathcal M}$ such that the support of $\pi$ is a subset of $\{1,
\ldots, L\}$. It can easily be shown that for any $L\geq 1$, the minimizer of $\Psi_{n}(\pi)$ over $\pi \in {\mathcal M}^{L}$ exists and is unique. We denote this minimizer by  $\widehat{\pi}^{L}$, and for any $L \geq \Mpt+1$, 
we calculate $\widehat{\pi}^{L}$  using the support 
reduction algorithm that was proposed by~\cite{GJW08}. 

Let us define the following notation. Let $\nu, \mu$ be two measures in ${\mathcal M}$. The derivative of
$\Psi_{n}$ in the direction $\nu$ calculated in $\mu$ is defined as
follows:
$$\left[D_{\nu}(\Psi_{n})\right](\mu)  =  \lim_{\varepsilon \downarrow 0}
\frac{1}{\varepsilon} \left(\Psi_{n}(\mu + \varepsilon\nu) -
  \Psi_{n}(\mu)\right),$$
  for all $\mu$ and $\nu$ such that $\Psi_{n}(\mu)$ and
  $\Psi_{n}(\nu)$ are finite.  
It can be written as 
\begin{equation}
 \left[D_{\nu}(\Psi_{n})\right](\mu)  =  \sum_{j\geq 1} \nu_{j} \left[d_{j}(\Psi_{n})\right](\mu) \label{eqD}
\end{equation}
where 
\begin{eqnarray*}
 \left[d_{j}(\Psi_{n})\right](\mu) & = & \lim_{\varepsilon \downarrow 0}
\frac{1}{\varepsilon} \left(\Psi_{n}(\mu + \varepsilon \delta_{j}) -
  \Psi_{n}(\mu)\right) \\
  &  =  &\sum_{l=0}^{j-1} T_{j}(l)
 \left(\sum_{j'\geq l+1} \mu_{j'}T_{j'}(l) - \widetilde{p}_{n}(l)\right).
\end{eqnarray*}

\paragraph{The algorithm for calculating $\widehat{\pi}^{L}$ for a
  fixed $L$ is described as follows:}
\begin{enumerate}
\item Initialisation
~\\
Let $S=\{L\}$ and choose  the
  measure $\pi^{L}$, such that 
\begin{eqnarray*}
\pi^{L}_{j} & =& 0 \mbox{ for } 1\leq j\leq L-1 \\
 \pi^{L}_{L} & = & \arg\min_{\pi\in\R}\sum_{i=0}^{L-1} \left(
  \widetilde{p}_{n}(i) - \pi T_{L}(i) \right)^{2}.
\end{eqnarray*} 
\item Optimisation over ${\mathcal M}^{L}$
~
\begin{description}
\item[Step 1:] For $1 \leq j \leq L $ calculate the quantities
  $\left[d_{j}(\Psi_{n})\right](\pi^{L})$.  If all  are  non negative, then
  set $\widehat{\pi}^{L} = \pi^{L}$, and the optimisation over
  ${\mathcal M}^{L}$ is achieved.  If not,  choose $j$ such that
  $\left[d_{j}(\Psi_{n})\right](\pi^{L}) 
  <0$, and set $S' = S + \{j\}$. For example, one can take $j$ as the
  minimizer of $\left[d_{j}(\Psi_{n})\right](\pi^{L})$. Go to step 2.
\item[Step 2:]  Let $\pi^{\star}_{S'}$ be the minimizer of
  $\Psi_{n}(\pi)$ over all measures $\pi$ such that ${\rm
      Supp}(\pi) \subset S'$. Two cases must be considered:
\begin{enumerate}
\item
If for all $l \in S'$, $\pi^{\star}_{S', l} \geq 0$, then set
  $\pi^{L} = \pi^{\star}_{S'}$, $S=S'$ and return to Step 1.
\item If not, let $l$ be defined as follows:
\begin{equation*}
l = {\rm arg}\min_{j'}
\left\{
  \varepsilon_{j'} = \frac{\pi^{L}_{j'}}
    {\pi^{L}_{j'}-\pi^{\star}_{S,j'}} \mbox{ for } j' \mbox{ such that } 
\pi^{\star}_{S,j'} < \pi^{L}_{j'}
\right\}.
\end{equation*}
Set  $S'=S + \{j\} - \{l\}$ and return to Step 2.
\end{enumerate}
\end{description}
%\item Updating $L$
%~ \\
%\com{Once the optimisation over ${\mathcal M}^{L}$ is stopped, we have at hand a measure $\hat\pi^L$}. If $\sum_{j=1}^{L} \widehat{\pi}_{j}^{L} =1$, then \com{set $\widehat{L}=L$ and $\hat\pi_n=\hat\pi^{\hat L}$, and the algorithm
%   is stopped.} 
   %$\widehat{L}=L$,  for all $0 \leq i \leq \widehat{L}$,  $\widehat{p}_{n}(i) =  \sum_{j=i+1}^{\widehat{L}}\widehat{\pi}_{j}^{\widehat{L}} T_{j}(i)$, and the algorithm is stopped. 
%   If not, take \com{$\pi^{L+1}=\widehat{\pi}^{L}$, $L=L+1$,} and go
%   to Step 1.
\end{enumerate}

\begin{theo}\label{theo: Algo}
The estimator $\widehat{\pi}^{L}$ given by the algorithm
  described above minimizes $\Psi_{n}(\pi)$ over $\pi \in {\cal M}^{L}$.
\end{theo}

Then, thanks to the following theorem, we are able to calculate 
a convenient $L$.

\begin{theo}\label{theo:AlgoF}
Let $L \geq \widetilde{s}_{n} + 1$. If $\widehat{\pi}^{L}$ is a
probability measure, then $\widehat{\pi}^{L} = \widehat{\pi}_{n}$.
\end{theo}

One possibility is to carry out the optimisation over ${\mathcal
  M}^{L}$ for increasing values of $L$ until the
condition $\sum_{j\geq 1}
\widehat{\pi}^{L}_{j} = 1$ is satisfied. As the support of
$\widehat{\pi}_{n}$ is finite, the condition will be fulfilled in a
finite number of steps.

%%%%%%%%%%%%%%%%%%%%%%%%%%%%%%%%%%%%%%%%%%%%%%%%%%%%%%%%%%%%%%%%%%%%%%
\section{Simulation study}
\label{simul.st}
%%%%%%%%%%%%%%%%%%%%%%%%%%%%%%%%%%%%%%%%%%%%%%%%%%%%%%%%%%%%%%%%%%%%%%

%%%%%%%%%%%%%%%%%%%%%%%%%%%%%%%%%%%%%%%%%%%%%%%%%%%%%%%%%%%%%%%%%%%%%%
\subsection{Simulation design}

We designed a simulation study to assess the quality of the
constrained estimator $\widehat{p}_n$ and to compare it with the
unconstrained estimator $\widetilde{p}_n$.

We considered two shapes for the distribution $p_0$: the geometric
$\cal G(\gamma)$ ($\gamma = .9, .5, .1$), the support of which is
infinite, and the pure triangular distribution $T_j$ ($j = 20, 5,
2$). For each distribution, we considered three sample sizes: $n = 10,
100$ and $1000$. 
We also considered the Poisson distribution with mean
$\lambda$, which is convex as long as $\lambda$ is smaller that
$\lambda^* = 2 - \sqrt{2} \simeq .59$. We considered $\lambda = .59$,
$.8$ and $1$.  For each simulation configuration, $1000$ random
samples were generated. The
simulation were carried out with {\tt R  (www.r-project.org)}, using
functions available at the following web-site \verb+http://w3.jouy.inra.fr/unites/miaj/public/perso/SylvieHuet_en.html+.

\subsection{Global fit}

\begin{figure}
  \includegraphics[angle=90,height=6cm, width=12cm]{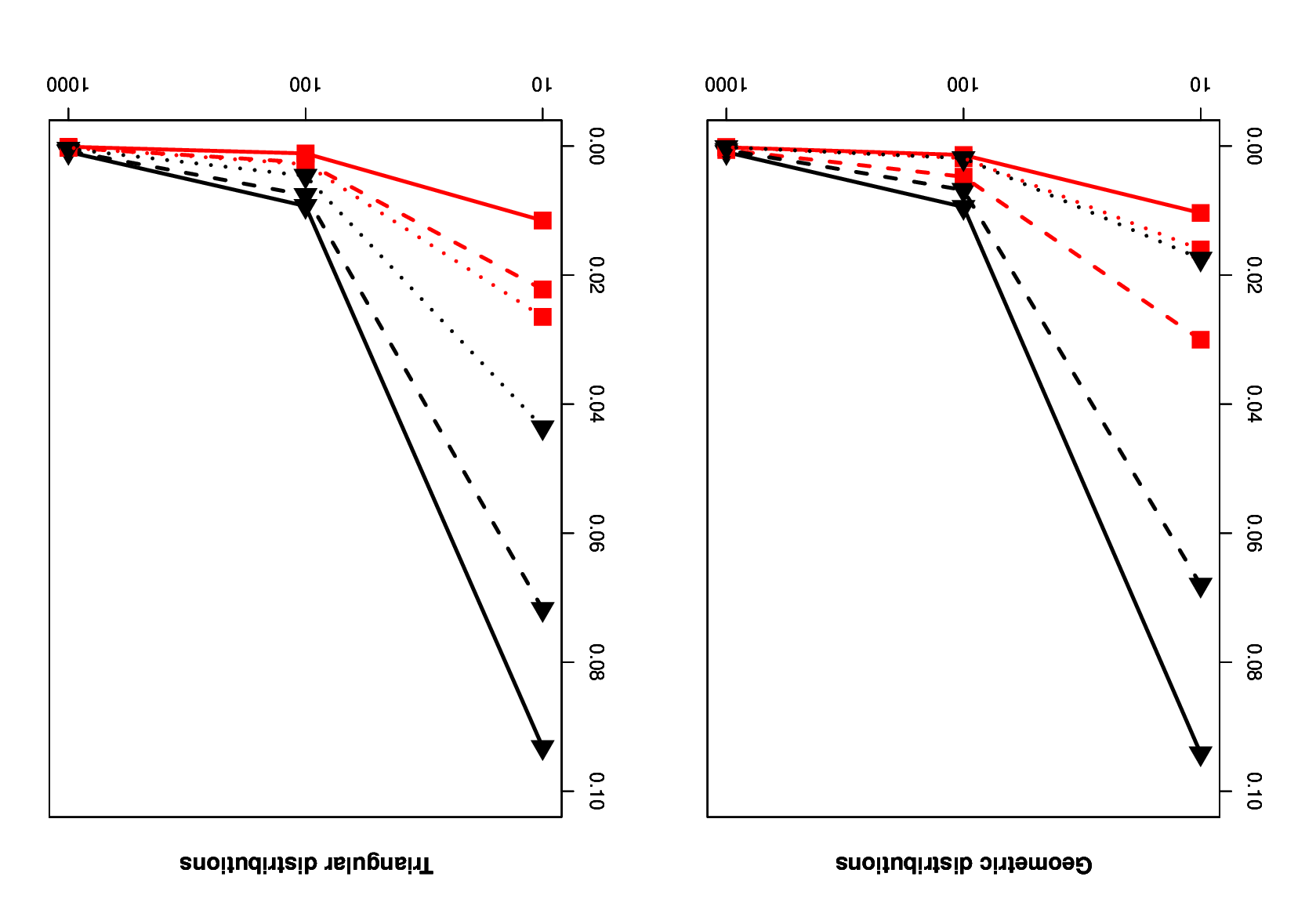} \\
  \includegraphics[angle=90,height=6cm, width=12cm]{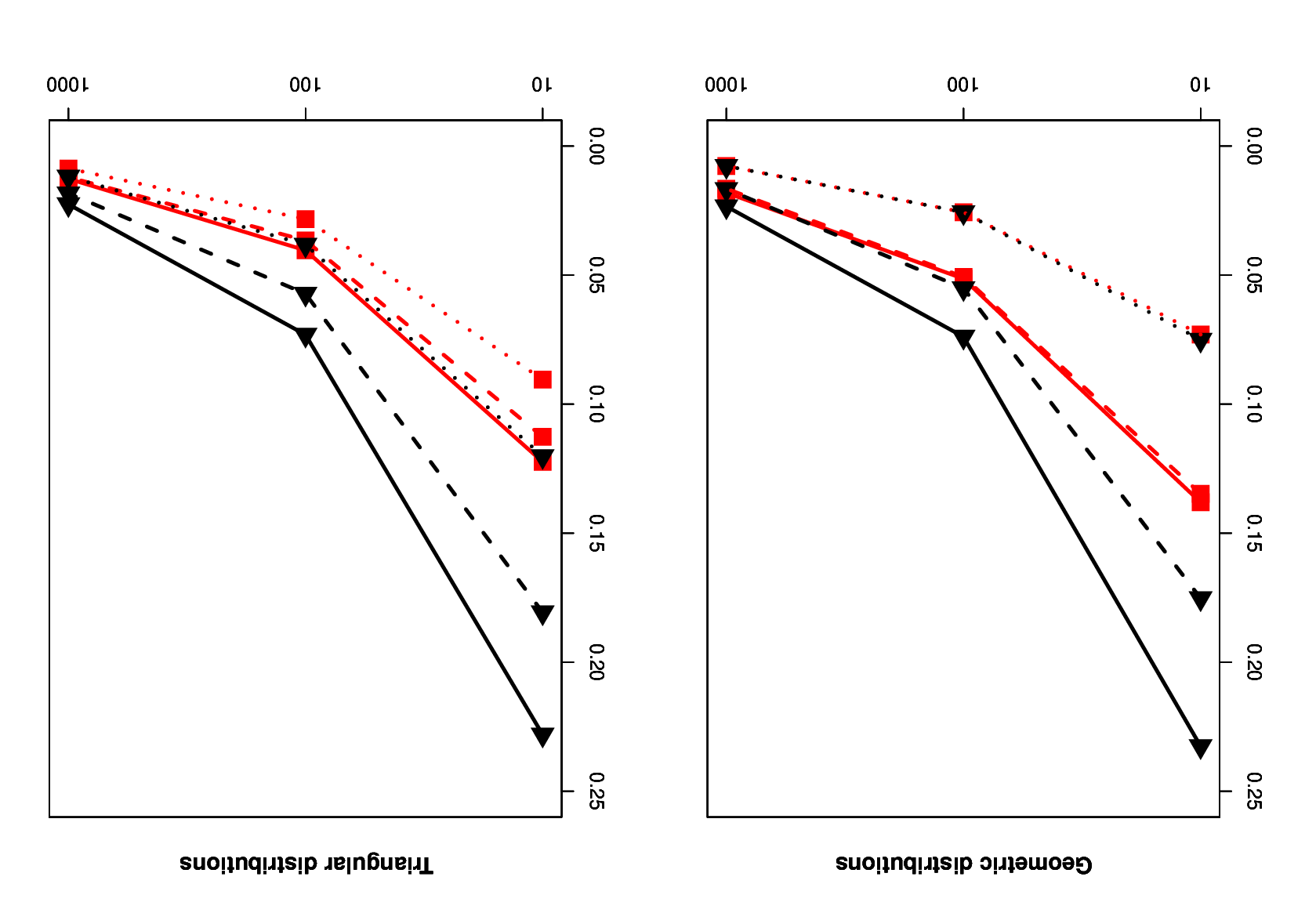}
  \caption{{\bf $\ell_2$-loss.} Empirical risk as a function of the
    sample size $n$. Top: $\ell_2(\cdot, p_0)$, bottom: $K(\cdot,
    p_0)$. Black: $\widetilde{p}_n$, \textcolor{red}{red}:
    $\widehat{p}_n$.  Solid ({\bf --}): $\gamma=1$ or $j = 20$, dashed
    (-\,-): $\gamma=.5$ or $j = 5$, dotted ($\cdots$): $\gamma=.9$ or
    $j =2$. \label{Fig:Loss} }
\end{figure}

We first compared the fit of the estimated distribution $\widehat{p}_n$
and $\widetilde{p}_n$ to the entire distribution $p_0$. To this aim, for
each simulated sample, we computed the $\ell_2$-loss for $\widehat{p}_n$
$$ 
\ell_2(\widehat{p}_n, p_0) = \sum_i [ \widehat{p}_n(i)  - p_0(i) ]^2,
$$ 
and likewise for $\widetilde{p}_n$. The expected $\ell_2$-loss
is estimated by the mean calculated on the basis of $1000$ simulations
and the results are displayed in Figure \ref{Fig:Loss}.

As expected from Theorem \ref{theo: empir/constraint}, the constrained
estimator $\widehat{p}_n$ outperforms the empirical estimator in all
configurations in terms of $\ell_2$-loss. The difference is larger in
the triangular case because of the existence of a region where $p_0$
is linear.  The empirical estimator $\widetilde{p}_n$ gets better and
closer to $\widehat{p}_n$ as the true distribution $p_0$ becomes more
convex, i.e., for $\gamma=.9$ or $j=2$. Note that the fit of the
unconstrained estimator improves when the true distribution gets
more convex.

These results are theoretically grounded by Theorem \ref{theo:
  empir/constraint} for the $\ell_2$-loss, but we also considered the
Kolomogorov loss:
$$
K(\widehat{p}, p_0) = \sup_i |\widehat{P}_n(i)  - P_0(i)|,
$$
where $P_0$ is the true cumulative distribution function (cdf) and
$\widehat{P}_n$ is the constrained cdf. The Kolmogorov loss of the
empirical cdf $\widetilde{P}_n$ was calculated in the same way.  As
shown on Figure \ref{Fig:Loss} (bottom), the behavior of the
Kolmogorov loss is similar to that of the $\ell_2$-loss. The same behavior
was observed for the Hellinger loss:
$$
\frac12 \sum_i \left(\sqrt{\widehat{p}_n(i)} - \sqrt{p_0(i)}\right)^2
$$
and the total variation loss:
$$
\frac1{2} \sum_i |\widehat{p}_n(i) - p_0(i)|.
$$
(results not shown). We thus observed that the constrained
estimator $\widehat{p}_n$ outperforms the empirical estimator for all
considered losses. 

%%%%%%%%%%%%%%%%%%%%%%%%%%%%%%%%%%%%%%%%%%%%%%%%%%%%%%%%%%%%%%%%%%%%%%
\subsection{Some characteristics of interest}
In this section, we consider the estimation of some characteristics of
the distribution, namely the variance, the entropy and the probability
at $0$. For each of these characteristics, denoted $\theta(p)$, we
measured the performance in terms of relative standard error:
$$
{\sqrt{\mathbb{E}\left(\theta(\widehat{p}_n) -
      \theta(p_0)\right)^2}} \left/ {\theta(p_0)}\right.. 
$$
% We focus on the bias of the estimators. 
The expectation was estimated by the mean over 1,000 simulations.

\begin{figure} 
  \includegraphics[angle=90,height=6cm, width=12cm]{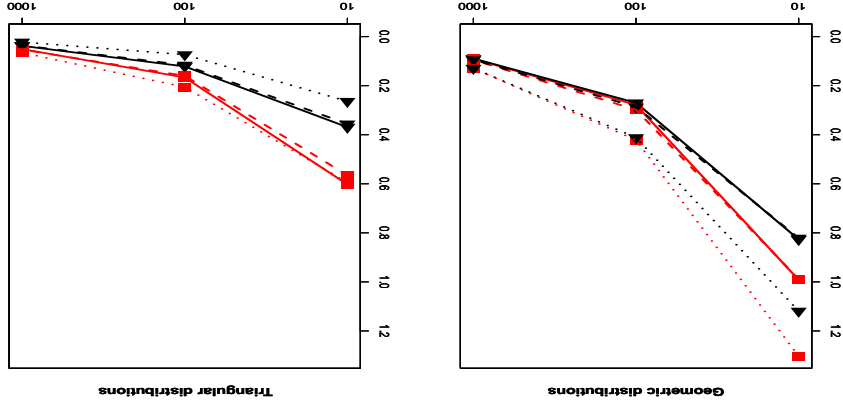}
  \caption{{\bf Variance.} Relative standard error of the variance as
    a function of the sample size $n$. Same legend as Figure
    \ref{Fig:Loss}. \label{Fig:Var}}
\end{figure}

As shown in Section \ref{sectionLSE}, the means of the empirical and
constrained distributions are equal, whereas the variance of the
constrained distribution is larger than the variance of the empirical one. Denoting by
$\mu_k$ the centered moment of order $k$ of $p_{0}$, the mean and variance of the
empirical variance are respectively
$$ 
\frac{n-1}n \mu_2 
\qquad
\text{and}
\qquad
\frac{n-1}{n^3}  \left((n-1)\mu_4 - (n-3) \mu_2^2\right).
$$
Figure \ref{Fig:Var} shows that the relative standard error of the
constrained estimator is smaller than that of the empirical
one. Hence, the constrained variance turns out to be more accurate.

\begin{figure}   
  \centering
  \includegraphics[angle=90,height=6cm,width=12cm]{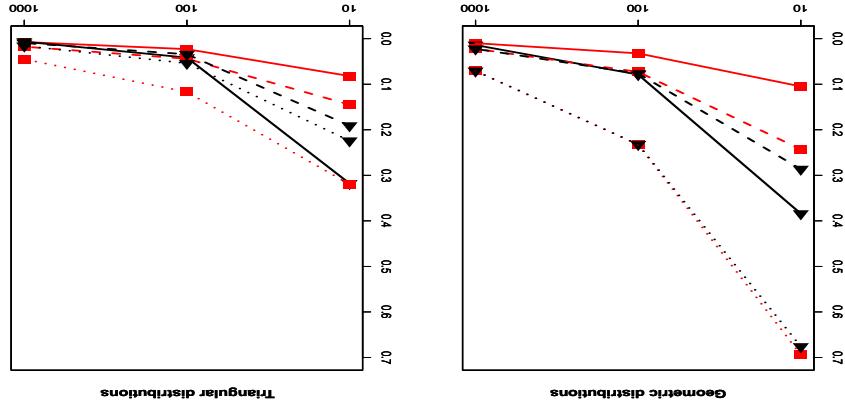} 
  \caption{{\bf Entropy.}  Relative standard error of the estimated
    entropy estimators as a function of the sample size $n$. Same
    legend as Figure \ref{Fig:Loss}. \label{Fig:Entropy} }
\end{figure}

We also investigated the estimation of the entropy 
$$ 
H(p) = -\sum_{i \geq 0}p(i)\log p(i),
$$
which is often used in ecology as a diversity
index. As shown in Figure \ref{Fig:Entropy},
  $H(\widehat p_n)$ is a better estimate of the true entropy than
  $H(\widetilde p_n)$, in most situations; the difference between
  the two estimators vanishes when the true distribution becomes more
  convex.
The worst performance of $H(\widehat p_n)$ are
  obtained when the true distribution is $T_2$. Note that this
  distribution is a special case since more than half of the
  estimation errors consist in adding a component $T_j$ ($j > 2$) in
  the mixture \eqref{eq: mixture k}, which result in an increase of
  the entropy.

\begin{figure}
  \includegraphics[height=6cm, width=12cm]{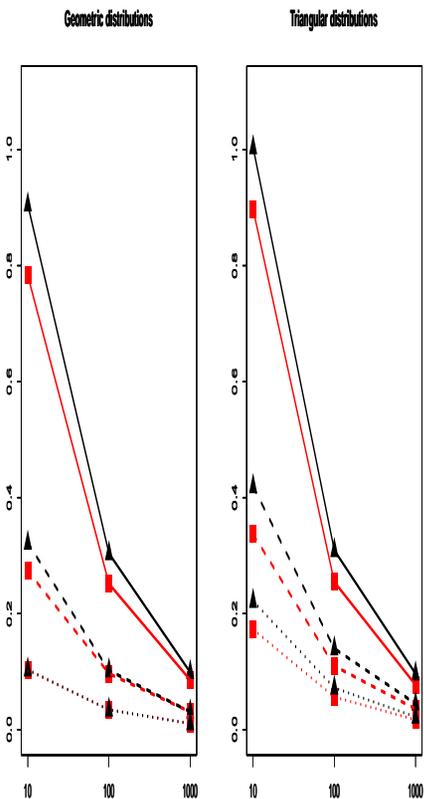}
  \caption{{\bf Probability mass in 0.} Relative standard error of the
    estimated probability mass in zero as a function of the sample
    size $n$. Same legend as Figure \ref{Fig:Loss}. \label{Fig:P0}}
\end{figure}

We then considered the estimation of the probability mass
$p(0)$. Theorem \ref{moments.th} showed that the constrained
estimator $\widehat{p}_n(0)$ is greater than or equal to the empirical
estimator $\widetilde{p}_n(0)$, which is known to be unbiased.
However, Figure \ref{Fig:P0} shows that the constrained estimator
$\widehat{p}_n$ still provides a more accurate estimate of $p_0(0)$
than $\widetilde{p}_n$.
  
For all these characteristics, the constrained distribution provides
better estimates than the empirical distribution, provided that the
true distribution is indeed convex.

%%%%%%%%%%%%%%%%%%%%%%%%%%%%%%%%%%%%%%%%%%%%%%%%%%%%%%%%%%%%%%%%%%%%%%
\subsection{Robustness to non-convexity}

We finally studied the robustness of the constrained estimator to non-convexity. As an example, we considered the Poisson distribution with
mean $\lambda$, which is convex as long as $\lambda$ is smaller that
$\lambda^* = 2 - \sqrt{2} \simeq .59$. We studied how
$\widetilde{p}_n$ and $\widehat{p}_n$ behave, in terms of $\ell_2$-loss, when $\lambda$ exceeds $\lambda^*$.

\begin{figure}
  \includegraphics[height=6cm, width=12cm]{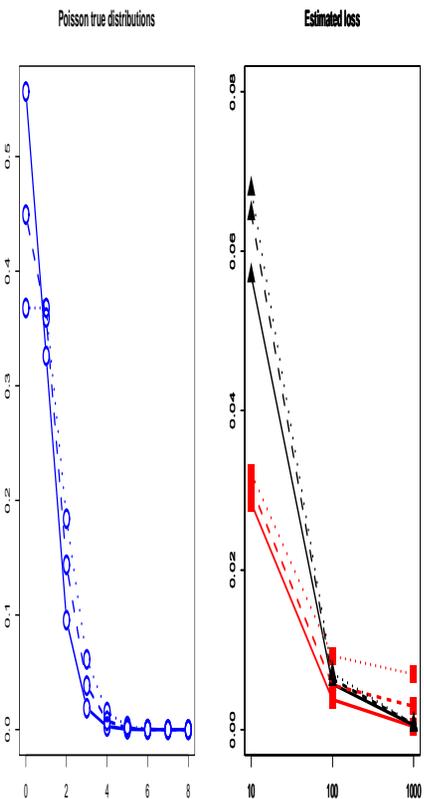}
  \caption{Left: Three different Poisson distributions.
    Solid ({\bf --}):$\lambda=\lambda^*$, dashed (-\,-):$\lambda=.8$,
    dotted ($\cdots$):$\lambda=1$. Right: empirical $\ell_2$-loss as a function
    of $n$. Black: $\widetilde{p}_n$, \textcolor{red}{red}:
    $\widehat{p}_n$. \label{Fig:Pois}}
\end{figure}

The left panel of Figure \ref{Fig:Pois} displays the Poisson
distributions with respective means $\lambda^*$, .8 and 1. 
Figure \ref{Fig:Pois} (right) shows that the $\ell_2$-loss of the
constrained estimator increases with $\lambda$. However for small
sample sizes, $\widehat{p}_n$ still provides a better fit than
$\widetilde{p}_n$, at least for $\lambda \leq 1$. The performance of
$\widehat{p}_n$ is dramatically altered when the sample size becomes
large and the convexity assumption is strongly violated.

\section{Proofs} \label{Proofs.st}

\subsection{Proof of Theorem~\ref{theo: LSE}}

In order to prove Theorem  \ref{theo: LSE}, we first prove in the
following lemma that the minimizer of $Q_n$ over ${\cal K}$ exists and
is unique, where ${\cal K}$ is the set defined in Section \ref{intro.st}. Then, after some intermediate results, we prove in
Lemma \ref{lem: LSE} below that the minimizer of $Q_n$ over ${\cal K}$
belongs to ${\cal C}$. Since ${\cal C}\subset{\cal K},$ Theorem
\ref{theo: LSE} follows from Lemma \ref{lem: min K} combined to Lemma
\ref{lem: LSE}. 

\paragraph{Notation}
We denote by $N_n$ the number of distinct values of the
$X_i$'s and by $X_{(1)},\dots,X_{(N_n)}$ these distinct values
rearranged in increasing order, i.e., such that
$X_{(1)}<\dots<X_{(N_n)}$. We set  $\mpt=X_{(1)}$ and
$\Mpt=X_{(N_n)}$. 

In the
case $\Mpt=0$ i.e., $\tilde p_n(0)=1$ and $\tilde p_n(i)=0$ for all
$i\geq 1$,  the proof of Theorem~\ref{theo: LSE} is
straightforward. Thus, in the sequel, we restrict ourselves
to the case $\Mpt\geq 1$.

 \begin{lem}\label{lem: min K}
There exists a unique $\hat p_n\in{\cal K}$ such that
\begin{equation}\label{eq: min K}
 Q_n(\hat p_n)=\inf_{p\in{\cal K}}Q_n(p).
\end{equation}
Moreover, $\hat p_n$ has a finite support.
\end{lem}

\paragraph{Proof.}
For proving the existence and uniqueness of $\widehat{p}_{n}$, we have
to prove the following preliminary results, where $q$ denotes a
candidate to be a minimizer of $Q_n$ over ${\cal K}$. 
\begin{itemize}
\item[(i)] There exists $c_1=c_1(\omega)<\infty$ that does not depend on $q$ such that $q\leq c_1$.
\item[(ii)] We have $q=\qt$ where
\begin{equation}\label{eq: tilde q}
 \qt(i)= \begin{cases}q(i) &\text{  for all } i \in\{0,\dots,\Mpt\}\\
\max \{q(\Mpt) + (q(\Mpt)-q(\Mpt-1))(i-\Mpt)\,,\,0\}&\text{  for all } i \geq \Mpt.
  \end{cases}
\end{equation}
\end{itemize}

Therefore, minimizing $Q_n$ over ${\cal K}$ amounts to minimizing $Q_n$
over the set of functions $q\in{\cal K}$ such that $q\leq c_1$,
$Q_n(q)\leq Q_n(T_1)$, and $q=\qt$. But for all $q\in{\cal K}$ such that $q=\qt$, we
have 
\begin{equation}\label{eq: Qn(q)} 
Q_n(q)=\frac{1}{2}\sum_{i=0}^{\Mpt}q^2(i)+\frac{1}{2}\sum_{i\geq 1}\left(\max\{q(\Mpt)+i(q(\Mpt)-q(\Mpt-1),0\}\right)^2-\sum_{i\geq 0}q(i)\tilde p_n(i) 
\end{equation}
and therefore, this amounts to minimizing
$$\Qt_n(t)=\frac{1}{2}\sum_{i=0}^{\Mpt}t^2(i)+\frac{1}{2}\sum_{i\geq 1}\left(\max\{t(\Mpt)+i(t(\Mpt)-t(\Mpt-1),0\}\right)^2-\sum_{i\geq 0}t(i)\tilde p_n(i)$$
over the set $K$ of non-increasing convex functions $t:\{0,\dots,\Mpt\}\to[0,\infty)$ such that $t(0)\leq c_1$ and $\Qt_n(t)\leq Q_n(T_1)$.
The set $K$ is compact and $\Qt_n$ is continuous and strictly convex on $K$, so there exists a unique minimizer of $\Qt_n$ over $K$. This proves that there exists a unique minimizer of $Q_n$ over ${\cal K}$. 

It remains to prove results (i) and (ii).

\paragraph{Proof of {\rm (i)}.}

It is easy to see that  for all $p \in {\cal K}$, 
\begin{equation*}
Q_{n} (p) \geq \frac{1}{2} p^{2}(\mpt) - p(\mpt)
\end{equation*}
using that $p$ is non-increasing. This lower bound tends to infinity
as $p(\mpt)\to\infty$.  But, if we consider $T_{1}$ the measure that
puts the mass 1 in 0, we have  $Q_n(q)\leq
Q_n(T_1)<\infty$, so there exists $c<\infty$ such that $q(\mpt)<c$. Now,
$Q_n(T_1)\geq Q_n(q)\geq q^2(0)/2-q(\mpt)$ and therefore, there exists
$c_1<\infty$ such that $q(0)\leq c_1$, which means that $q\leq c_1$. 

\paragraph{Proof of {\rm (ii)}.}

By convexity we must have $\qt(i)\leq q(i)$ for all $i\geq \Mpt$ and
therefore, $$Q_{n}(q)-Q_{n}(\qt)=\sum_{i> \Mpt}(q^2(i)-\qt^2(i))/2\geq
0$$  
with a strict inequality in the case $q\neq\qt$. This proves that any
candidate $q$ to be a minimizer of $Q_n$ over ${\cal K}$ should
satisfy $q=\qt$.

Let us now prove  that the support of $\widehat{p}_{n}$ is finite. 
In the case $\hat p_n(\Mpt)=0$, it is clear that $\hat p_n$ has a
finite support included in $\{0,\dots,\Mpt-1\}$. Consider the case $\hat
p_n(\Mpt)>0$.  Let us first remark that $\hat p_n(\Mpt-1)>\hat p_n(\Mpt)$, since otherwise, we
would have $\hat p_n(i)=\hat p_n(\Mpt)$ for all $i\geq \Mpt$ so that
$Q_n(\hat p_n)=\infty$. Then define $\qt$ as in (\ref{eq: tilde q})
where $q$ is 
replaced by $\hat p_n$. 
From the proof of Lemma \ref{lem: min K}, we
know that $\hat p_n = \qt$ which has finite support as soon as   $\hat
p_n(\Mpt-1)>\hat p_n(\Mpt)$.\cqfd

\paragraph{~}

The following lemma provides a precise characterization of $\hat p_n$. It is the counterpart, in the discrete case, of Lemma 2.2 in \cite{GJW01} for the continuous case. For every $p\in {\cal K},$ we define
\begin{equation}\label{def: FH}
F_p(j)=\sum_{i=0}^j p(i)\text{ and }H_p(j)=\sum_{i=0}^j F_p(i)
\end{equation}
for all integers $j\geq 0$, and $F_p(j)=H_p(j)=0$ for all integers $j<0$. Thus, $F_p$ is a distribution function in the case $p\in{\cal C}$.

\begin{lem}\label{lem: charact}
Let $\hat p_n$ be the unique function in ${\cal K}$ that satisfies (\ref{eq: min K}). For all $l\geq 1$ we have
\begin{equation}\label{eq:  charact}
H_{\hat p_n}(l-1)\geq H_{\tilde p_n}(l-1)
\end{equation}
with an equality if $\hat p_n$ has a change of slope at point $l$, i.e., if
$$\hat p_n(l)-\hat p_n(l-1)<\hat p_n(l+1)-\hat p_n(l).$$
Conversely, if $p\in{\cal K}$ satisfies $H_{p}(l-1)\geq H_{\tilde p_n}(l-1)$ for all $l\geq 1$ with an equality if $p(l)-p(l-1)<p(l+1)-p(l)$, then $p=\hat p_n$. 
\end{lem}

\paragraph{Proof.} First, note that $\tilde p_n$ has a finite support by
definition, and Lemma \ref{lem: affine} ensures that $\hat p_n$ has a
finite support as well. Thus, all the sums involved in the proof are
well-defined and finite.  
For every $\eps>0$ and $l\geq 1$, define $q_{\eps l}$ by $q_{\eps l}(i)=\hat p_n(i)$ for all $i\geq l$ and
$$q_{\eps l}(i)=\hat p_n(i)+\eps (l-i) $$
for all $i\in\{ 0,\dots, l\}$. Thus, $q_{\eps l}$ is the sum of convex functions, which implies that $ q_{\eps l}\in{\cal K}$ for all $\eps,l$. Since $\hat p_n$ minimizes $Q_n$ over ${\cal K}$, we have  $Q_n(q_{\eps l})\geq Q_n(\hat p_n)$ for all $\eps,l$ and therefore,
$$\liminf_{\eps\downarrow0}\frac{1}{\eps}\big(Q_n(q_{\eps l})-Q_n(\hat p_n)\big)\geq 0$$
for all $l\geq 1$. This simplifies to
$$\sum_{i= 0}^{l-1}\hat p_n(i)(l-i)\geq \sum_{i= 0}^{l-1}\tilde p_n(i)(l-i)$$
for all $l\geq 1$ and can be rewritten as
$$\sum_{j=0}^{l-1}\sum_{i= 0}^{j}\hat p_n(i)\geq \sum_{j=0}^{l-1}\sum_{i= 1}^{j}\tilde p_n(i)$$
for all $l\geq 1$, which is precisely (\ref{eq:  charact}). To prove
the equality case, note that $(1+\eps)\hat p_n\in{\cal K}$ for all
$\eps>-1$. Therefore, for all $\eps>-1$ we have 
$$Q_n\big((1+\eps)\hat p_n\big)\geq Q_n(\hat p_n).$$
Distinguishing the cases $\eps>0$ and $\eps<0$ we obtain
$$\liminf_{\eps\downarrow0}\frac{1}{\eps}\big(Q_n((1+\eps)\hat p_n)-Q_n(\hat p_n)\big)\geq 0$$
and
$$\limsup_{\eps\uparrow0}\frac{1}{\eps}\big(Q_n((1+\eps)\hat p_n)-Q_n(\hat p_n)\big)\leq 0.$$
Both limits are equal, so their common value is equal to zero, which can be written as 
$$\sum_{i\geq 0}\hat p_n(i)\big(\hat p_n(i)-\tilde p_n(i)\big)=0.$$
Now, noticing that $p(i)=F_p(i)-F_p(i-1)$ for all $p\in{\cal K}$ and $i\in\N$, we arrive at
\[\begin{split}
 0=&\sum_{i\geq 0}\hat p_n(i)\Big(F_{\hat p_n}(i)-F_{\hat p_n}(i-1)-F_{\tilde p_n}(i)+F_{\tilde p_n}(i-1)\Big)\\
=&\sum_{i\geq 0}\hat p_n(i)\Big(F_{\hat p_n}(i)-F_{\tilde p_n}(i)\Big)-\sum_{i\geq 1}\hat p_n(i)\Big(F_{\hat p_n}(i-1)-F_{\tilde p_n}(i-1)\Big).\\
\end{split}\]
Rearranging the indices, we have
$$\sum_{i\geq 1}\hat p_n(i)\Big(F_{\hat p_n}(i-1)-F_{\tilde p_n}(i-1)\Big)=\sum_{i\geq 0}\hat p_n(i+1)\Big(F_{\hat p_n}(i)-F_{\tilde p_n}(i)\Big),$$
whence
$$0=\sum_{i\geq 0}\Big(\hat p_n(i)-\hat p_n(i+1)\Big)\Big(F_{\hat p_n}(i)-F_{\tilde p_n}(i)\Big).$$
Now, we notice that $F_p(i)=H_p(i)-H_p(i-1)$ for all $p\in{\cal K}$ and $i\in\N$. A similar change of indices as above then yields
$$0=\sum_{i\geq 0}\Big((\hat p_n(i)-\hat p_n(i+1))-(\hat p_n(i+1)-\hat p_n(i+2))\Big)\Big(H_{\hat p_n}(i)-H_{\tilde p_n}(i)\Big).$$
It follows from (\ref{eq:  charact}) that $H_{\hat p_n}(i)\geq H_{\tilde p_n}(i)$ for all $i\geq 0$, and we have 
$$\hat p_n(i+1)-\hat p_n(i)\leq\hat p_n(i+2)-\hat p_n(i+1)$$
by convexity of $\hat p_n$. A sum of non-negative numbers is equal to zero if and only if these numbers are all equal to zero, so we conclude that
$$\Big((\hat p_n(i)-\hat p_n(i+1))-(\hat p_n(i+1)-\hat
p_n(i+2))\Big)\Big(H_{\hat p_n}(i)-H_{\tilde p_n}(i)\Big)=0$$ 
for all $i\geq 0$. Hence, $H_{\hat p_n}(i)=H_{\tilde p_n}(i)$ for all
$i\geq 0$ that satisfy  
$$\hat p_n(i+1)-\hat p_n(i)<\hat p_n(i+2)-\hat p_n(i+1).$$
Setting $l=i+1$, this means that we have an equality in (\ref{eq:  charact}) if $\hat p_n$ has a change of slope at point $l$.

Conversely, consider $p\in{\cal K}$ such
that $H_{p}(i)\geq H_{\tilde p_n}(i)$ for all $i\geq 0$ with an
equality if $p(i+1)-p(i)<p(i+2)-p(i+1)$. Then we have 
\begin{equation}\label{eq: 0=} 
0=\sum_{i\geq 0}\big(p(i)-2p(i+1)+p(i+2)\big)\big(H_{p}(i)-H_{\tilde p_n}(i)\big)
\end{equation}
and $p$ has a finite support.  
To see this, argue by contradiction and assume for a while that the support of $p$ is not finite. In such a case, there exists an increasing sequence $(u_l)_{l\in\N}$ such that $u_l$ tends to infinity as $l\to\infty$ and $p$ has changes of slope at every point $u_l+1$, $l\in\N$. This implies that 
$$H_{p}(u_l)-H_p(u_{l-1})= H_{\tilde p_n}(u_l)-H_{\tilde p_n}(u_{l-1})$$ 
for all $l\geq 1$. Using that $F_p$ is non-increasing and that $\tilde p_n$ has a finite support, we obtain
$$F_p(u_{l-1})\leq\frac{1}{u_{l}-u_{l-1}}\left(H_{p}(u_l)-H_p(u_{l-1})\right)= F_{\tilde p_n}(u_l)=\sum_{i\geq 0}\tilde p_n(i)$$
for all large enough $l$ and similarly,
 $$F_p(u_{l})\geq\sum_{i\geq 0}\tilde p_n(i)$$
for all large enough $l$. Therefore, 
$$F_p(u_l)=F_p(u_{l-1})=\sum_{i\geq 0}\tilde p_n(i)$$ 
for all large enough $l$, which means that $p(i)=0$ for all large
enough $i$. This is in contradiction with the assumption that the
support of $p$ is not finite, which proves that the support of $p$ is
finite.

Now, let $q\in{\cal K}$ be any candidate to be a minimizer of $Q_n$
over ${\cal K}$. We know, see the proof of Lemma \ref{lem: min K},
that $Q_n(q)\leq Q_n(T_1)$, and $q=\qt$, where $\qt$ is defined by
(\ref{eq: tilde q}). In particular, $q$ satisfies (\ref{eq: Qn(q)})
which implies that  $q$ has a finite
support. Thus, we can write
\begin{eqnarray}\label{eq: charac egal}
 Q_n(q)-Q_n(p)&=&\frac{1}{2}\sum_{i\geq 0}\Big(q^2(i)-p^2(i)-2\tilde p_n(i)\big( q(i)-p(i)\big)\Big)\notag\\
&=&\frac{1}{2}\sum_{i\geq 0}\Big(q^2(i)-p^2(i)+2\big(p(i)-\tilde p_n(i)\big)\big( q(i)-p(i)\big)-2p(i)\big( q(i)-p(i)\big)\Big)\notag\\
&=&\frac{1}{2}\sum_{i\geq 0}\big(q(i)-p(i)\big)^2+\sum_{i\geq 0}\big(p(i)-\tilde p_n(i)\big)\big( q(i)-p(i)\big)\notag\\
&\geq&\sum_{i\geq 0}\big(p(i)-\tilde p_n(i)\big)\big( q(i)-p(i)\big).
\end{eqnarray}
Using that both $q$ and $p-\tilde p_n$ have a finite support
and rearranging the indices as above, we show that
\[\begin{split}
   \sum_{i\geq 0}&\big(p(i)-\tilde p_n(i)\big)\big( q(i)-p(i)\big)\\
&=\sum_{i\geq 0}\Big(\big(q(i)-p(i)\big)-2\big(q(i+1)-p(i+1)\big)+\big(q(i+2)-p(i+2)\big)\Big)\big(H_p(i)-H_{\tilde p_n}(i)\big).
  \end{split}\]
Combining this with (\ref{eq: 0=}) and (\ref{eq: charac egal}) yields
$$Q_n(q)-Q_n(p)\geq\sum_{i\geq 0}\big(q(i)-2q(i+1)+q(i+2)\big)\big(H_p(i)-H_{\tilde p_n}(i)\big).$$
The right-hand side is non-negative since $H_{p}(i)\geq H_{\tilde p_n}(i)$ for all $i\geq 0$ and $q$ is convex over $\N$, so we conclude that $Q_n(q)\geq Q_n(p)$ for all candidates $q\in{\cal K}$. This means that $p$ minimizes $Q_n$ over ${\cal K}$. \cqfd\\

We are now in a position to prove that $\hat p_n$ is a probability mass function, i.e., $\hat p_n\in{\cal C}.$ 

\begin{lem}\label{lem: LSE}
Let $\hat p_n$ be the unique function in ${\cal K}$ that satisfies (\ref{eq: min K}).
We have 
\begin{equation}\label{eq: Ftilde=Fhat}
 F_{\tilde p_n}(\Mph + 1)=F_{\hat p_n}(\Mph + 1),
\end{equation}
$\Mph \geq \Mpt$ and $\hat p_n\in{\cal C}.$
\end{lem}

\paragraph{Proof.} 

Let us first prove by contradiction that $\Mph$ is well-defined.  Let $k= 1+\min_{j}\left\{\widetilde{p}_{n}(j)
  \neq 0\right\}$. It is easy to verify that there exists a
strictly positive $a$ such that $Q_{n}(a T_{k}) < 0$. As $Q_{n}(0)=0$,
 $\widehat{p}_{n}$ cannot be
identically zero and $\Mph$ is well-defined.

By definition of $\Mph$, $\hat p_n$ has a change of slope at point
$\Mph + 1$, so it follows from Lemma \ref{lem: charact} that
\begin{equation}\label{eq: egal s-1}
\sum_{j=0}^{\Mph}F_{\hat p_n}(j)= \sum_{j=0}^{\Mph}F_{\tilde p_n}(j). 
\end{equation}
Using Lemma \ref{lem: charact} again we obtain 
$$\sum_{j=0}^{\Mph + 1}F_{\hat p_n}(j)\geq \sum_{j=0}^{\Mph+1}F_{\tilde p_n}(j),$$
which, combined with (\ref{eq: egal s-1}) shows that $F_{\hat
  p_n}(\Mph+1)\geq F_{\tilde p_n}(\Mph+1)$. 

Let us first consider the case where $\Mph\geq 1$. We have 
$$\sum_{j=0}^{\Mph-1}F_{\hat p_n}(j)\geq \sum_{j=0}^{\Mph-1}F_{\tilde p_n}(j)$$
which, combined with (\ref{eq: egal s-1}) shows that $F_{\hat
  p_n}(\Mph)\leq F_{\tilde p_n}(\Mph)$. But $\hat p_n(\Mph+1)=0$ by
definition of $\Mph$, so we also have  $F_{\hat
  p_n}(\Mph+1)=F_{\hat p_n}(\Mph)$ and therefore,  
$$F_{\tilde p_n}(\Mph)\geq F_{\hat p_n}(\Mph+1)\geq F_{\tilde
  p_n}(\Mph+1).$$  
 By definition, $F_{\tilde p_n}$ is
  non-decreasing, so we conclude that (\ref{eq: Ftilde=Fhat}) holds. 

Consider now the case $\Mph=0$.  We have $\tilde
  p_n(1)=0$: otherwise, we could modify $\hat p_n$ to a $q\in{\cal
  K}$ such that $q (0)=\hat p_n(0)$,  $0<q(1)\leq \tilde p_n(1)$ and
  $q(i)=0$ for all $i>1$, which is a contradiction since for such a
  $q$ we have $Q_n(q)<Q_n(\hat p_n)$. Moreover, in the case
  $\Mph=0$, we have $\hat p_n(0)=\tilde p_n(0)$: otherwise, we
  could modify $\hat p_n$ to a $q\in{\cal K}$ such that $q
  (0)=\tilde p_n(0)$  and $q(i)=0$ for all $i>0$ which is a
  contradiction since for such a $q$ we have $Q_n(q)<Q_n(\hat
  p_n)$. Hence,  
$$F_{\hat p_n}(1)=\hat p_n(0)=\tilde p_n(0)=F_{\tilde p_n}(1),$$ 
which completes the proof of (\ref{eq: Ftilde=Fhat}).

For the purpose of proving that $\Mph \geq \Mpt$, we argue by
contradiction. Assume for a while that $\Mph=\Mpt-1$. This means that
$\hat p_n(i)=0$ for all $i\geq \Mpt$ and $\hat p_n(\Mpt-1)>0$. In this
case, we can modify $\hat p_n$ to a $q\in{\cal K}$ such that $q
(i)=\hat p_n(i)$ for all $i<\Mpt$,  $0<q(\Mpt)\leq \tilde p_n(\Mpt),$
and $q(i)=0$ for all $i>\Mpt$. Then we have 
\begin{eqnarray*}
2\big(Q_n(q)-Q_n(\hat p_n)\big)&=&\sum_{i\geq 0}\big (q(i)-\tilde p_n(i)\big)^2-\sum_{i\geq 0}\big (\hat p_n(i)-\tilde p_n(i)\big)^2 \\
&=&\big (q(\Mpt)-\tilde p_n(\Mpt)\big)^2-\big (\tilde p_n(\Mpt)\big)^2\\
&<&0.
\end{eqnarray*}
This is a contradiction since $\hat p_n$ minimizes $Q_n$ and
therefore, $\Mph\neq \Mpt-1.$ Assume now that $\Mph<\Mpt-1$. Then,
$F_{\tilde p_n}(\Mph+1)<1$, so (\ref{eq: Ftilde=Fhat}) yields 
$$F_{\hat p_n}(j)=F_{\hat p_n}(\Mph+1)<1$$
for all $j\geq \Mph+1$. Therefore, for all $l>\Mpt$ we have
$$\sum_{j=0}^{l-1}\big(F_{\hat p_n}(j)-F_{\tilde
  p_n}(j)\big)=\sum_{j=0}^{\Mpt-1}\big(F_{\hat p_n}(j)-F_{\tilde
  p_n}(j)\big)+(l-\Mpt)\big (F_{\hat p_n}(\Mph+1)-1\big),$$ 
which tends to $-\infty$ as $l\to\infty$. This is a contradiction since from Lemma \ref{lem: charact}, this has to remain non-negative for all $l$. We conclude that $\Mph\geq\Mpt$. Combining this with (\ref{eq: Ftilde=Fhat}) yields
$$F_{\hat p_n}(\Mph+1)= F_{\tilde p_n}(\Mph+1)=1.$$
This proves that $\hat p_n$ is a probability mass function and
completes the proof of the lemma. \cqfd 

\subsection{Proof of Theorem~\ref{theo: empir/constraint}}

Let us begin with the following lemma that gathers together a number of properties of the
minimizer $\hat p_n$. These
properties compare to those of the constrained least squares
estimator of a convex density function over $[0,\infty)$, see
\cite{GJW01}: in this case  the constrained LSE has a bounded support, is piecewise linear,
has no changes of slope at the observation points, and has at most one
change of slope between two consecutive observation points. In the
 discrete case, the constrained LSE is also piecewise linear with
bounded support. However, due to the fact that $\N$ is a discrete set,
the constrained LSE can have changes of slopes at the observation
points and can have two changes of slopes between two consecutive
observations.

\begin{lem}\label{lem: affine}
The unique function $\hat p_n\in{\cal K}$ that satisfies (\ref{eq: min
  K}) has the following properties:
$\hat p_n$ is linear on the
  interval $\{0,\dots, X_{(1)}+1\}$ and also on
  $\{\Mpt -1,\dots,\Mph\}$; in the case where $N_n$, the number of distinct values of the
$X_i$'s, is greater or equal to 2, it has at most two changes of slopes on $\{X_{(j)},\dots,X_{(j+1)}\}$ for any given $j=1,\dots, N_n-1$, and in the case where it has two changes of slopes on this set, these changes occur at consecutive points in $\N$. 
\end{lem}

\paragraph{Proof.} 

We know, from the proof of Lemma~\ref{lem: min K}, that $\hat p_n =
\qt$, where $\qt$ is defined as in  (\ref{eq: tilde q})
where $q$ is 
replaced by $\hat p_n$. It follows that $\hat p_n$ is linear on
$\{\Mpt -1,\dots,\Mph\}$ in the case $\Mph \geq \Mpt$. Consider an
arbitrary candidate $p$  to be a minimizer of $Q_n$ over ${\cal K}$,
fix $j\in\{1,\dots,N_n-1\}$, and define the functions $p_l$ and $p_r$
over $\N$ as follows:  $p_l(i)=p(i)$ for all $i\leq X_{(j)}+1$ and all
$i\geq X_{(j+1)}$ and $p_l$ is linear over
$\{X_{(j)},...,X_{(j+1)}-1\}$, whereas $p_r(i)=p(i)$ for all $i\leq
X_{(j)}$ and all $i\geq X_{(j+1)}-1$ and $p_r$ is linear over
$\{X_{(j)}+1,...,X_{(j+1)}\}$. Setting $q(i)=\max\{p_l(i),p_r(i)\}$
for all $i\in\N$, we obtain that $q\in{\cal K}$ is piecewise linear
over $\{X_{(j)},...,X_{(j+1)}\}$ with at most two changes of slopes
over this interval and in case it has two changes of slopes, these
changes occur at consecutive points. We have $q(X_{(j)})=p(X_{(j)})$
for all $j$, and $q\leq p$ by convexity of $p$. Since $\tilde
p_n(i)>0$ if and only if $i=X_{(j)}$ for some $j$, this implies that
$Q_n(q)\leq Q_n(p)$ with a strict inequality if $p\neq q$. Therefore,
$p$ could be a minimizer of $Q_n$ only if $p=q$. This implies that the
minimizer $\hat p_n$ is piecewise linear over
$\{X_{(j)},\dots,X_{(j+1)}\}$ with at most two changes of slopes over
this interval. A similar argument shows that $\hat p_n$ is linear over
the interval $\{0,\dots, X_{(1)}+1\}$. \cqfd\\ 

\subsubsection*{Proof of Equation~(\ref{eq.theo2})}

We prove that (\ref{eq.theo2}) holds with $p_0$ replaced by any $q \in {\mathcal K}$ that belongs to $l_2$, i.e., that satisfies $\sum_{j\geq 
  0}q^2(j)<\infty$.  Since
$p_{0}$ belongs to $l_1$ as a probability mass function and  $l_2\subset l_1$,  $p_0$ also belongs to $l_2$, so (\ref{eq.theo2}) with $p_0$ replaced by any $q \in {\mathcal K}$ that belongs to $l_2$ is a slightly more general result than (\ref{eq.theo2}).  

Consider an arbitrary $q \in {\mathcal K}$ satisfying
$\sum_{j\geq 
  0}q^2(j)<\infty$.
We have
\[%\begin{split}
\sum_{j\geq 0}\big(q(j)-\tilde p_n(j)\big)^2\geq \sum_{j\geq 0}\big(q(j)-\hat p_n(j)\big)^2+2\sum_{j\geq 0}\big(\hat p_n(j)-\tilde p_n(j)\big)\big(q(j)-\hat p_n(j)\big)
  %\end{split}
\]
with a strict inequality in the case where $\tilde p_n$ is non-convex
since in that case, $\tilde p_n\neq \hat p_n$. Thus, in order to prove
that (\ref{eq.theo2}) holds with $p_0$ replaced by $q$, it suffices to prove that
\begin{equation}\label{eq: empir/constraint}
 \sum_{j\geq 0}\big(\hat p_n(j)-\tilde p_n(j)\big)\big(q(j)-\hat p_n(j)\big)\geq0.
\end{equation}
According to Lemma~\ref{lem: affine}, there exist integers
$c_0<\dots<c_m$ such that $c_0=0$, $c_m=\Mph+1$, $\hat p_n$ is linear
over the interval $\{c_{i-1},\dots,c_{i}\}$ and has a change of slope
at point $c_i$, for all $i=1,\dots,m$. It follows from Theorem
\ref{theo: LSE} that $\Mph \geq \Mpt$,  so $\tilde p_n(j)=\hat
p_n(j)=0$ for all $j\geq \Mph+1$ and the sum in (\ref{eq:
  empir/constraint}) can be split as follows: 
\begin{equation}\label{eq: split}
\sum_{j\geq 0}\big(\hat p_n(j)-\tilde p_n(j)\big)\big(q(j)-\hat
p_n(j)\big)= \sum_{i=1}^{m} \sum_{j=c_{i-1}}^{c_i-1}\big(\hat p_n(j)-\tilde p_n(j)\big)f(j) 
\end{equation}
where $f(j)=q(j)-\hat p_n(j)$ for all $j\geq 0$.  For all $i=1,\dots,m$ we have 
\[\begin{split}
 \sum_{j=c_{i-1}}^{c_i-1}&\big(\hat p_n(j)-\tilde p_n(j)\big)f(j)\\
 &=\sum_{j=c_{i-1}}^{c_i-1}\left[\big(F_{\hat p_n}(j)-F_{\tilde p_n}(j)\big)-\big(F_{\hat p_n}(j-1)-F_{\tilde p_n}(j-1)\big)\right]f(j)\\
&=\sum_{j=c_{i-1}}^{c_i-1}\big(F_{\hat p_n}(j)-F_{\tilde p_n}(j)\big)f(j)-\sum_{j=c_{i-1}-1}^{c_i-2}\big(F_{\hat p_n}(j)-F_{\tilde p_n}(j)\big)f(j+1)\\
&=\sum_{j=c_{i-1}}^{c_i-1}\big(F_{\hat p_n}(j)-F_{\tilde p_n}(j)\big)\left(f(j)-f(j+1)\right)\\
&\quad  +\big(F_{\hat p_n}(c_i-1)-F_{\tilde p_n}(c_i-1)\big)f(c_i)\\
&\quad\quad  -\big(F_{\hat p_n}(c_{i-1}-1)-F_{\tilde p_n}(c_{i-1}-1)\big)f(c_{i-1}),
\end{split}\]
where $F_{\hat p_n}$ and $F_{\tilde p_n}$ are defined in (\ref{def: FH}).
By definition, $F_{\tilde p_n}(j)=F_{\hat p_n}(j)=0$ for all $j<c_0$, so summing up over $i$ yields
\begin{eqnarray*}
 \sum_{i=1}^{m}\sum_{j=c_{i-1}}^{c_i-1}\big(\hat p_n(j)-\tilde
 p_n(j)\big)f(j)&=&\sum_{i=1}^{m}\sum_{j=c_{i-1}}^{c_i-1}\big(F_{\hat
   p_n}(j)-F_{\tilde p_n}(j)\big)\left(f(j)-f(j+1)\right)\\  
& & +\big(F_{\hat p_n}(c_m-1)-F_{\tilde p_n}(c_m-1)\big)f(c_m),
\end{eqnarray*}
where we recall that $c_m=\Mph+1$. Now, it follows from the definition
of $\Mph$ that $\hat p_n(\Mph+1)=0$ and we also have $\tilde
p_n(\Mph+1)=0$ since $\Mph\geq\Mpt$, see Theorem~\ref{theo: LSE}. Thanks to
(\ref{eq: Ftilde=Fhat}), we conclude that $F_{\tilde
  p_n}(\Mph)=F_{\hat p_n}(\Mph)$. Therefore,  (\ref{eq:
  split}) combined with the preceding display yields 
\[\begin{split}\sum_{j\geq 0}&\big(\hat p_n(j)-\tilde p_n(j)\big)\big(q(j)-\hat p_n(j)\big)\\ &=\sum_{i=1}^{m}\sum_{j=c_{i-1}}^{c_i-1}\big(F_{\hat p_n}(j)-F_{\tilde p_n}(j)\big)\left(f(j)-f(j+1)\right).\end{split}\]
Now, $H_{\tilde p_n}(j)=H_{\hat p_n}(j)=0$ for all $j<c_0$ and $F_p(j)=H_p(j)-H_p(j-1)$ for $p=\tilde p_n,\hat p_n$ and all $j$, so we can repeat the same arguments as above to obtain 
\[\begin{split}
 \sum_{i=1}^m&\sum_{j=c_{i-1}}^{c_i-1}\big(F_{\hat p_n}(j)-F_{\tilde p_n}(j)\big)\left(f(j)-f(j+1)\right)\\
&=\sum_{i=1}^m\sum_{j=c_{i-1}}^{c_i-1}\big(H_{\hat p_n}(j)-H_{\tilde p_n}(j)\big)\left(f(j)-2f(j+1)+f(j+2)\right)\\
&\quad +\big(H_{\hat p_n}(c_m-1)-H_{\tilde p_n}(c_m-1)\big)(f(c_m)-f(c_m+1)).
\end{split}\]
Since $\hat p_n$ has a change of slope at each $c_i$, we deduce from Lemma \ref{lem: charact} that $H_{\hat p_n}(c_i-1)=H_{\tilde p_n}(c_i-1)$ for all $i=1,\dots,m$ and we arrive at
\begin{equation}\label{eq: splitH}\begin{split}
 \sum_{j\geq 0}&\big(\hat p_n(j)-\tilde p_n(j)\big)\big(q(j)-\hat p_n(j)\big)\\
 &=\sum_{i}\sum_{j=c_{i-1}}^{c_i-2}\big(H_{\hat p_n}(j)-H_{\tilde p_n}(j)\big)\left(f(j)-2f(j+1)+f(j+2)\right),\end{split}
\end{equation}
where the first sum on the right-hand side is taken over those
$i=1,\dots,m$ such that $c_{i-1}\leq {c_i-2}$. For such an $i$, $f$ is
convex over the interval $\{c_{i-1},\dots,c_i\}$ as a sum of a convex
function and a linear function (recall that by definition of the
$c_i$'s, $\hat p_n$ is linear over such an interval). Therefore we get
$$f(j)-2f(j+1)+f(j+2)\geq 0$$ 
for all $j=c_{i-1},\dots,c_i-2$, see (\ref{def:
  convex(slope)}). Moreover, it follows from Lemma \ref{lem: charact}
that $H_{\hat p_n}\geq H_{\tilde p_n}$, which leads to 
$$\big(H_{\hat p_n}(j)-H_{\tilde p_n}(j)\big)\left(f(j)-2f(j+1)+f(j+2)\right)\geq 0$$
for all $j=c_{i-1},\dots,c_i-2$. Combining this with (\ref{eq:
  splitH}) yields (\ref{eq: empir/constraint}) and completes the proof
of the first part of the theorem.

\subsubsection*{Proof of Equations~(\ref{eq: tilde p non-convex}) and~(\ref{liminf.eq})}

It suffices to prove (\ref{eq: tilde p non-convex}) since the second
assertion follows from (\ref{eq: tilde p non-convex}) and~\ref{eq.theo2}. To prove (\ref{eq: tilde p
  non-convex}), note that 
$$\P\Big(\tilde p_n\text{ is non-convex}\Big)\geq \P\left(\tilde p_n(i)-2\tilde p_n(i+1)+\tilde p_n(i+2)<0\right)$$
and that by assumption, we have
$p_0(i)-2p_0(i+1)+p_0(i+2)=0$. Therefore, we have the following inequality:
\[\begin{split}
&\P\Big(\tilde p_n\text{ is non-convex}\Big)\\
&\quad \geq
 \P\left(\sqrt n\Big[ (\tilde p_n(i)-p_0(i))-2(\tilde p_n(i+1)-p_0(i+1))+(\tilde p_n(i+2)-p_0(i+2))\Big]<0\right).  
  \end{split}\] 
  From the central limit theorem, the random variable
$$\sqrt n\Big[ (\tilde p_n(i)-p_0(i))-2(\tilde p_n(i+1)-p_0(i+1))+(\tilde p_n(i+2)-p_0(i+2))\Big]$$
converges, as $n\to\infty$, to a centered Gaussian variable $X$ with a non-degenerate variance and therefore,
$$\liminf_{n\to\infty}\P\left(\tilde p_n\text{is non-convex}\right)\geq \P(X\leq 0).$$
The lemma follows since $\P(X\leq 0)=1/2$. \cqfd

\subsection{Proof of Theorem~\ref{moments.th}}

Let us first note  that for any positive  concave
function $q$ defined on $\N$, such that $q(\widehat{s}_{n})>0$ and $q(i) =0$ for all $i >
\widehat{s}_{n}$, the function $\widehat{p}_{n} - \varepsilon q$
 belongs to ${\cal K}$ as soon as $\varepsilon \leq
 \widehat{p}_{n}(\widehat{s}_{n})/q(\widehat{s}_{n})$. 

Besides, thanks to Theorem~\ref{theo: LSE}, we know that $\widehat{p}_{n}
=\mbox{ Argmin}_{f \in {\cal K }} Q_n(f)$. Therefore for all $q$ defined as
above, 
\begin{eqnarray*}
 0 &\leq &\lim_{\varepsilon \searrow 0} \frac{Q_n(\hat p_n - \varepsilon q)-Q_n(\hat p_n)}{\varepsilon}\\
   & =   &\sum_{i=0}^{\hat s_n }(\tilde p_n(i) -\hat p_n(i))q(i).
\end{eqnarray*}
Let $u\geq 1$, $0\leq a \leq \widehat{s}_{n}$, and take 
\begin{equation*}
q(i) = \left(1-
  \left(\frac{|i-a|}{\widehat{s}_{n}+1-a}\right)^{u}\right), \mbox{
  for } 1 \leq i
\leq \widehat{s}_{n}, \end{equation*}
 and  $q(i)=0$ for  $i > \widehat{s}_{n}$. 
Then we get the inequality in Equation~(\ref{moments.ineq}). 

The proof of  $\sum_{i=1}^{\widetilde{s}_{n}} i
\widetilde{p}_{n}(i)=\sum_{i=1}^{\widehat{s}_{n}} i
\widehat{p}_{n}(i)$ follows from the fact that the function
$\widehat{p}_{n} + \varepsilon q$ belongs to ${\cal K}$ for 
\begin{equation*}
q(i) = \left(1 - \frac{i}{\widehat{s}_{n}+1}\right)  \mbox{
  for } 1 \leq i
\leq \widehat{s}_{n}, \mbox{ and } q(i)=0, \mbox{ for } i > \widehat{s}_{n}.
\end{equation*}
It remains to prove that $\hat p_{n}(0)\geq \tilde p_{n}(0)$. Argue by
contradiction and assume that $\hat p_{n}(0)< \tilde p_{n}(0)$. Define
$q(0)=\tilde p_{n}(0)$ and $q(i)=\hat p_{n}(i)$ for all $i\geq
1$. Then, $q\in {\cal K}$ since $\hat p_{n}$ is convex and $q(0)\geq
\hat p_{n}(0)$, and we have $Q_{n}(q)<Q_{n}(\hat p_{n})$. This is a
contradiction since $\hat p_{n}$ minimizes $Q_{n}$ over ${\cal K}$,
see Theorem \ref{theo: LSE}. This completes the proof of the
theorem.

\subsection{Proof of Theorem~\ref{theo: mixture k}}

Assume $f\in{\cal K}$ and consider the function $\pi$ defined by
(\ref{eq: pi/p}). The function $\pi$ takes non-negative values since
$f$ is convex, see (\ref{def: convex(slope)}). Therefore  $\pi$
belongs to ${\cal M}$. Moreover, for all $i\in\N$ we have 
\begin{eqnarray*}
 \sum_{j\geq i+1} \pi_jT_j(i)&=&\sum_{j\geq i+1} \big( f(j+1)+f(j-1)-2f(j)\big)(j-i)\\
&=&\sum_{j\geq i+1} \sum_{k=1}^{j-i}\big( f(j+1)+f(j-1)-2f(j)\big).\\
\end{eqnarray*}
Since all terms in the sum are non-negative and $\lim_{i\to\infty}f(i)=0$, we can write
\begin{eqnarray*}
\sum_{j\geq i+1} \pi_jT_j(i)&=&\sum_{k\geq 1} \sum_{j\geq i+k}\big( f(j+1)+f(j-1)-2f(j)\big),\\
&=&\sum_{k\geq 1}\big( f(i+k-1)-f(i+k)\big)\\
&=&f(i)
\end{eqnarray*}
for all $i\in\N$. Therefore, $\pi\in{\cal M}$ satisfies (\ref{eq:
  mixture k}). Conversely, every $f:\N\to [0,\infty)$ satisfying
(\ref{eq: mixture k}) for some $\pi\in{\cal M}$ is clearly convex, so
we obtain the first assertion of the theorem. To prove the second and
the third assertions, we assume that $f\in{\cal K}$. So, in view of the
preceding result, we know that $f$ satisfies (\ref{eq: mixture k}) for
some $\pi\in{\cal M}$. Thus, we have 
\begin{eqnarray*}
f(i-1)-f(i)&=&\sum_{j\geq i}\pi_{j} \big (T_{j}(i-1)-T_{j}(i)\big)\\
&=&\sum_{j\geq i}\frac{2\pi_{j}}{j(j+1)}
\end{eqnarray*}
for all $i\geq 1$. By convexity of $f$ we conclude that
\begin{eqnarray*}
0\leq \big(f(i-1)-f(i)\big)-\big(f(i)-f(i+1)\big)=\frac{2\pi_{i}}{i(i+1)}
\end{eqnarray*}
for all $i\geq 1$, which implies that  $\pi$ is uniquely defined by (\ref{eq: pi/p}). Moreover,
$$\sum_{i\geq 0}f(i)=\sum_{i\geq 0}\sum_{j\geq i+1}\pi_jT_j(i)=\sum_{i\geq 0}\sum_{j\geq 1}\pi_jT_j(i)$$
since $T_j(i)=0$ for all $j\leq i$. This implies that
$$\sum_{i\geq 0}f(i)=\sum_{j\geq 1}\pi_j\big(\sum_{i\geq 0}T_j(i)\big)$$
where $\sum_{i\geq 0}T_j(i)=1$. This completes the proof of the theorem.
\cqfd

\subsection{Proof of Theorem~\ref{theo: Algo}} \label{proofAlgo.st}

The theorem is proved following the work of Groeneboom and
al.~\cite{GJW08}. It follows from Lemmas~\ref{lem12} and~\ref{lem13}
given below.

\begin{lem}\label{lem12}
Let  $\Mpt$ be the
maximum of the support
of $\widetilde{p}_{n}$ and $L \geq \Mpt+1$. Then we have the following
result:
$\widehat{\pi}^{L} = \arg\min_{\mu \in {\mathcal M}^{L}}
 \Psi_{n}(\mu)$ is equivalent to
\begin{equation}
 \left[d_{j}(\Psi_{n})\right](\widehat\pi^{L}) \geq 0 \; \forall 1
  \leq j \leq L, \mbox{ and } \left[d_{j}(\Psi_{n})\right](\widehat\pi^{L}) = 0
 \;  \forall j \in \mbox{Supp}(\widehat\pi^{L}) \label{P2.eq}
\end{equation}
\end{lem}

\paragraph{Proof.}
Let $\widehat\pi^{L} = \arg\min_{\mu \in {\mathcal M}^{L}}
 \Psi_{n}(\mu)$.

For all $1 \leq j \leq L$ and $\varepsilon >0$, $\widehat\pi^{L} +
\varepsilon \delta_{j} \in {\mathcal M}^{L}$, and
$\Psi_{n}(\widehat\pi^{L} + \varepsilon \delta_{j}) \geq
\Psi_{n}(\widehat\pi^{L})$. It follows that
$\left[d_{j}(\Psi_{n})\right](\widehat\pi^{L}) \geq 0$.

If $j \in \mbox{Supp}(\widehat\pi^{L})$, then for $\varepsilon > 0$
small enough, $\widehat\pi^{L} -\varepsilon \delta_{j} \in {\mathcal
  M}^{L}$, and $\Psi_{n}(\widehat\pi^{L} - \varepsilon \delta_{j}) \geq
\Psi_{n}(\widehat\pi^{L})$. It follows that
$-\left[d_{j}(\Psi_{n})\right](\widehat\pi^{L}) \geq 0$.

Conversely, assume that Equation~(\ref{P2.eq}) is satisfied, and
 take $\pi
 \in {\mathcal M}^{L}$. Then $\left[D_{\pi}(\Psi_{n})\right](\widehat\pi^{L})$
 is non negative and 
 $\left[D_{\widehat\pi^{L}}(\Psi_{n})\right](\widehat\pi^{L})=0$,
thanks to Equation~(\ref{eqD}). 

By convexity of $\Psi_{n}$, for $\varepsilon>0$
\begin{equation*}
\Psi_{n}(\pi)  - \Psi_{n}(\widehat\pi^{L}) \geq \frac{1}{\varepsilon} \left(
\Psi_{n}(\varepsilon \pi + (1-\varepsilon) \widehat\pi^{L}) - \Psi_{n}(\widehat\pi^{L})\right).
\end{equation*}
Taking the limit when $\varepsilon$ tends to 0, we get
\begin{eqnarray*}
 \Psi_{n}(\pi)  - \Psi_{n}(\widehat\pi^{L}) 
&\geq & 
\left[D_{\pi-\widehat\pi^{L}}(\Psi_{n})\right](\widehat\pi^{L}) \\
&=& \left[D_{\pi}(\Psi_{n})\right](\widehat\pi^{L})  \geq 0.
\end{eqnarray*}\cqfd

\begin{lem}\label{lem13}
Let us define the following quantities.
\begin{itemize}
\item Let $\pi=\sum_{i=1}^{L-1} a_{i} \delta_{j_{i}}$ be the minimizer
of $\Psi_{n}$ over the set of positive measures spanned by
$\left\{ \delta_{j_{i}},\; 1 \leq i \leq L-1\right\}$.
\item Let $j_{L}$ be an integer such that $j_L \neq j_i$ for all $i=1,
  \ldots, L-1$, and $\left[d_{j_{L}}(\Psi_{n})\right](\pi) <0$.
\item Let $\pi^{\star} 
= \sum_{i=1}^{L} b_{i} \delta_{j_{i}}$ be the minimizer of $\Psi_{n}$ over the set spanned by
$\left\{ \delta_{j_{i}}, \; 1 \leq i \leq L\right\}$.
\end{itemize}
Then
$b_{L} >0$, and there exists $\varepsilon>0$ such that $\pi +
\varepsilon (\pi^{\star} - \pi)$ is a  non negative measure, and
such that $\Psi_{n}(\pi +\varepsilon (\pi^{\star} - \pi) )< \Psi_{n}(\pi)$.
\end{lem}

\paragraph{Proof.}
 Following the same arguments as in the proof of Lemma~\ref{lem12}, we
 have $\left[d_{j_{i}}(\Psi_{n})\right](\pi) = 0$ for all $i=1,
 \ldots, L-1$. Then
\begin{equation*}
 \left[D_{\pi}(\Psi_{n})\right](\pi) = \sum_{i=1}^{L-1}
a_{i} \left[d_{j_{i}}(\Psi_{n})\right](\pi) = 0.
\end{equation*}

Moreover, we have
\begin{equation*}
 \Psi_{n}(\pi + \varepsilon \delta_{j_{L}}) = \Psi_{n}(\pi) +
 \frac{\varepsilon^{2}}{2} + \varepsilon \left[d_{j_{L}}(\Psi_{n})\right](\pi).\end{equation*}
Therefore, $\left[d_{j_{L}}(\Psi_{n})\right](\pi) <0$ implies that for
$\varepsilon >0$ small enough, 
\begin{equation*}
 \Psi_{n}(\pi + \varepsilon \delta_{j_{L}}) < \Psi_{n}(\pi).
\end{equation*}
This shows that $\pi^{\star} \neq \pi$.

By convexity of $\Psi_{n}$, we show that
\begin{eqnarray*}
 \lim_{\varepsilon \downarrow 0} 
\frac{\Psi_{n}((1-\varepsilon)\pi + \varepsilon \pi^{\star}) -
   \Psi_{n}(\pi) }{\varepsilon}  & \leq &  
\lim_{\varepsilon \downarrow 0}
 \frac{(1-\varepsilon)\Psi_{n}(\pi) + 
\varepsilon \Psi_{n}(\pi^{\star})-\Psi_{n}(\pi)}{\varepsilon}  \\
 &=&\Psi_{n}(\pi^{\star})-\Psi_{n}(\pi) <0.
\end{eqnarray*}
This shows that for $\varepsilon>0$ small enough, $\Psi_{n}(\pi + \varepsilon
(\pi^{\star}-\pi)) < \Psi_{n}(\pi)$.

Besides, we have
\begin{eqnarray*}
 \lim_{\varepsilon \downarrow 0} \frac{1}{\varepsilon} \left(
   \Psi_{n}((1-\varepsilon)\pi + \varepsilon \pi^{\star}) -
   \Psi_{n}(\pi)\right) & = & \left[D_{\pi^{\star} -
     \pi}(\Psi_{n})\right](\pi) \\
 & = & \left[D_{\pi^{\star}}(\Psi_{n})\right](\pi) \\
& = & \sum_{i=1}^{L} b_{i} \left[d_{j_{i}}(\Psi_{n})\right](\pi)\\
& = & b^{L}  \left[d_{j_{L}}(\Psi_{n})\right](\pi) .
\end{eqnarray*}

Because $\left[d_{j_{L}}(\Psi_{n})\right](\pi) <0$, $b_{L}$ is
positive.

It remains to show that there exists $\varepsilon >0$ such that, for
all $1 \leq i \leq L-1$, $a_{i} + \varepsilon (b_{i}-a_{i})$ is non
negative. This is clearly the case if $b_{i} \geq a_{i}$. If not, take
$\varepsilon \leq \min_{b_{i} <
  a_{i}}\left\{a_{i}/(a_{i}-b_{i})\right\}$. \cqfd

\subsection{Proof of Theorem~\ref{theo:AlgoF}} \label{proofAlgoF.st}

%Then, we have to show that $\widehat{L}=\Mph+1$. We know that the minimum of $\Psi_{n}$ over ${\mathcal M}$ is the minimum of $\Psi_{n}$ over the set of probability measures. Thus, starting from $L=\Mpt+1$, we know that $L < \Mph+1$ as long as $\sum_{i=1}^{L}\widehat{\pi}^{L} \neq 1$.  By construction, $\widehat{L}$ is the smallest $L$ (greater than $\Mpt+1$) such that $\sum_{i=1}^{L} \widehat{\pi}^{L} = 1$. Because, for all $\Mpt+1 \leq L \leq \widehat{L}$,  ${\mathcal   M}^{L} \subset {\mathcal  M}^{\widehat{L}}$, $\Psi_{n}(\widehat{\pi}^{\widehat{L}}) \leq \Psi_{n}(\widehat{\pi}^{L})$. It remains to show that the same inequality holds for all $L >\widehat{L}$. 

%\com

%This follows from Lemma~\ref{lem14} where
%it is proved that if for $L\geq \Mpt+1$, $\widehat{\pi}^{L}$ is a
%probability measure, then for all $L'> L$,
%$\widehat{\pi}^{L'}=\widehat{\pi}^{L}$.
Let us begin with the following lemma.
\begin{lem}\label{lem14}
If $\widehat{\pi}^{L} = \arg\min_{\mu \in {\mathcal M}^{L}}
 \Psi_{n}(\mu)$, then for all $j \geq 1$,
\begin{equation*}
 \left[d_{L+j}(\Psi_{n})\right](\widehat{\pi}^{L}) = b\left(\sum_{i=1}^{L} \widehat{\pi}^{L}_{i}
-1\right),
\end{equation*}
for some positive constant $b$ depending on $j$ and on the maximum of
the support of $\widehat{\pi}^{L}$.
\end{lem}

\paragraph{Proof.}
Let us consider two cases according to wether
$\widehat{\pi}^{L}_{L}$ equals 0 or not. 

Suppose that $\widehat{\pi}^{L}_{L} > 0$, and  write
\begin{eqnarray*}
 \left[d_{L+j}(\Psi_{n})\right]({\widehat\pi}^{L}) & =  & \sum_{l=1}^{L+j-1}
 T_{L+j}(l) \left( \sum_{j'=l+1}^{L} {\widehat\pi}^{L}_{j'} T_{j'}(l) -
 \widetilde{p}_{n}(l)\right) \\
 & =  & \sum_{l=1}^{L-1}
 T_{L+j}(l) \left( \sum_{j'=l+1}^{L}\widehat\pi^{L}_{j'} T_{j'}(l) -
 \widetilde{p}_{n}(l)\right).
\end{eqnarray*}
Because for $0 \leq l \leq L-1$, $T_{L+j}(l) = a T_{L}(l) + b $, for
constants $a$ and $b$ depending on $L$ and $j$, we get 
\begin{equation*}
  \left[d_{L+j}(\Psi_{n})\right](\widehat\pi^{L})  = a
  \left[d_{L}(\Psi_{n})\right](\widehat\pi^{L}) + b \left[\sum_{l=1}^{L-1}
    \sum_{j'=l+1}^{L} \widehat\pi_{j'}^{L} T_{j'}(l) -1\right].
\end{equation*}

Following Lemma~\ref{lem12}, $\left[d_{L}(\Psi_{n})\right](\widehat\pi^{L})
=0$, and we get 
\begin{equation*}
\left[d_{L+j}(\Psi_{n})\right](\widehat\pi^{L})  =  b \left( \sum_{i=1}^{L}\widehat\pi^{L}_{i} \sum_{l=0}^{j-1}T_{j}(l)
  -1 \right) = b  \left( \sum_{i=1}^{L}\widehat\pi^{L}_{i}-1 \right).
\end{equation*}

If $\widehat{\pi}^{L}_{L} =0$, then $\widehat{\pi}^{L} \in {\mathcal M
  }^{L_1}$ for some $L_{1} < L$. Thanks to Lemma~\ref{lem12}, we know
  that $\widehat{\pi}^{L}$ is the minimizer of $\Psi_{n}$ over 
${\mathcal  M}^{L_1}$. Then we can show that
  $\left[d_{L_{1}+j}(\Psi_{n})\right](\widehat\pi^{L}) =0 $ for all
  $j\geq 1$ exactly as we have done in the case $\widehat{\pi}^{L}_{L} > 0$.

\paragraph{~}

To conclude the proof of Theorem~\ref{theo:AlgoF}, note first that for all $L'\leq L$, we have ${\mathcal   M}^{L'} \subset
{\mathcal  M}^{L}$, which implies
\begin{equation}\label{eq: hatpihatL}\Psi_{n}(\widehat{\pi}^{L}) \leq \Psi_{n}(\widehat{\pi}^{L'}).
\end{equation}
Second, it follows from Lemmas~\ref{lem12} and~\ref{lem14} that if $\sum_{i=1}^{L} \widehat{\pi}^{L}_{i}=1$, then for all
$L'\geq L$, $\widehat{\pi}^{L'}=\widehat{\pi}^{L}$. Therefore
Equation~(\ref{eq: hatpihatL}) 
holds for all positive integers $L'$, which 
implies that $$\Psi_{n}(\widehat{\pi}^{L}) \leq \Psi_{n}(\pi)$$ for
all measures $\pi\in\cal M$ with a finite support. Therefore
$\widehat{\pi}^{L} = \widehat{\pi}_{n}$. \cqfd

\bibliographystyle{elsarticle-harv}
\bibliography{ConvDiscr}




%% \bibitem must have one of the following forms:
%%   \bibitem[Jones et al.(1990)]{key}...
%%   \bibitem[Jones et al.(1990)Jones, Baker, and Williams]{key}...
%%   \bibitem[Jones et al., 1990]{key}...
%%   \bibitem[\protect\citeauthoryear{Jones, Baker, and Williams}{Jones
%%       et al.}{1990}]{key}...
%%   \bibitem[\protect\citeauthoryear{Jones et al.}{1990}]{key}...
%%   \bibitem[\protect\astroncite{Jones et al.}{1990}]{key}...
%%   \bibitem[\protect\citename{Jones et al., }1990]{key}...
%%   \harvarditem[Jones et al.]{Jones, Baker, and Williams}{1990}{key}...
%%


% \end{thebibliography}

\end{document}